# CSwin2SR: Circular Swin2SR for Compressed Image Super-Resolution

Honggui Li
*College of Information Engineering*
Yanghzou University
Yangzhou, China
hgli@yzu.edu.cn

Maria Trocan
*LISITE Research Laboratory*
ISEP
Paris, France
maria.trocan@isep.fr

Mohamad Sawan
*College of Engineerig*
Westlake University
Hangzhou, China
sawan@westlake.edu.cn

Dimitri Galayko
*Laboratoire d'Informatique de Paris 6*
Sorbonne University
Paris, France
dimitri.galayko@sorbonne-universite.fr

*Abstract*—Closed-loop negative feedback mechanism is extensively utilized in automatic control systems and brings about extraordinary dynamic and static performance. In order to further improve the reconstruction capability of current methods of compressed image super-resolution, a circular Swin2SR (CSwin2SR) approach is proposed. The CSwin2SR contains a serial Swin2SR for initial super-resolution reestablishment and circular Swin2SR for enhanced super-resolution reestablishment. Simulated experimental results show that the proposed CSwin2SR dramatically outperforms the classical Swin2SR in the capacity of super-resolution recovery. On DIV2K valid dataset, the average increment of PSNR is greater than 0.18 dB and the related average increment of SSIM is greater than 0.01.

*Keywords—closed-loop, negative feedback, Swin2SR, CSwin2SR, compressed image, super-resolution*

## I. Introduction

With the exponential growth of realistic and virtual image data, it is necessary to compress the original image for the purpose of highly-efficient image storage, transmission, processing, analysis, understanding and some emergent applications, such as augment reality and metaverse. In order to gain a high ratio of image compression, lossy image compression methods are preferentially considered. However, lossy image compression will unavoidably lead to the distortion to the original image. Hence, there is a urgent demand for compressed image restoration, including compressed image denoising, deblurring, deblocking, inpainting, artifact removing and super-resolution (SR) [1].

Classical compressed image SR methods utilized an open-loop architecture to obtain a high-quality image from a low-quality image. Closed-loop negative feedback framework is broadly used in automatic control systems and attains outstanding static and dynamic performance [2]. For the sake of further lifting the capability of existing open-loop compressed image SR approaches, a closed-loop structure is introduced.

Shifted windows V2 transformer for compressed image super-resolution and restoration (Swin2SR) is one of the state-of-the-art algorithms for compressed image SR [3]. Swin2SR is also one of the top solutions at the advances in image manipulation challenge on SR of compressed image and video. This paper attempts to enhance the image recovery performance of Swin2SR by forming a closed-loop architecture. The new algorithm is named as circular Swin2SR (CSwin2SR). To the best of our knowledge, there are no previous work on closed-loop compressed image SR.

The main contributions of this paper are listed as follows: (1) a serial Swin2SR architecture is proposed by adding down-sampling and compression units into the traditional Swin2SR; (2) a circular Swin2SR architecture is proposed by building a closed-loop negative feedback mechanism; (3) dramatic performance improvement of CSwin2SR is gained compared with the conventional Swin2SR.

The remainder of this paper is arranged as follows. Section 2 is related work, section 3 is theoretical basis, section 4 is simulation experiments, and section 5 is final conclusions.

## II. Related Work

Compressed image SR is one of the challenge branches of image restoration [4]. It can be classified into two categories: model-based method and learning-based method. Model-based method wields hand-crafted priors and learning-based method wields data-driven priors [4, 5]. Currently, learning-based method is the mainstream direction. Learning-based method makes use of convolutional neural network (CNN) and vision transformer. Compared with classical CNN, emerging vision transformer has the advantage of long-range attention and achieves better image SR performance.

Shifted windows (Swin) transformer is a hierarchical general-purpose baseline for computer vision [6]. Swin transformer acquires greater efficiency by confining self-attention to non-overlapping local windows while permitting cross-window linking. SwinV2 transformer reaches excellent performance in image classification, object detection and semantic segmentation by scaling up capacity and resolution [7]. SwinV2 transformer puts forward some innovations: a residual post normalization algorithm and a scaled cosine attention algorithm to strengthen the stability of large vision transformer models; a log-spaced continuous position bias algorithm to effectively transfer vision transformer models at low-resolution images and windows to their high-resolution versions. Swin transformer image restoration (SwinIR) is the pioneering research on image SR via Swin transformer [8]. SwinIR consists of three subparts: shallow subnetwork, deep subnetwork and reconstruction subnetwork. Swin2SR is the upgradation of SwinIR through SwinV2 transformer [3].

Compressed image SR can also be divided into two categories: known-compression method and unknown-compression method. For the known-compression method, the image compression algorithm and its parameters are known. For instance, lossy JPEG image compression algorithm is employed and its parameter of compression quality is given. For unknown-compression method, the compression algorithm or its parameters is indeterminate. It is more difficult to resolve unknown-compression method than known-compression method. This paper focuses on the known-compression method. The proposed method fully takes advantage of the determinate compression algorithm and its parameters to establish a circular structure for upgrading the SR reconstruction quality.

It should be mentioned that local self-attention negative-feedback module has already been adopted for image SR [9]. However, this module does not build a global closed-loop architecture.

## III. Theory

### A. Improved Open-Loop Swin2SR

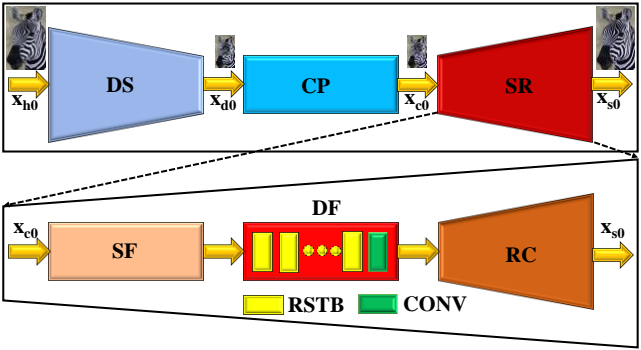

Fig. 1. Architecture of improved open-loop Swin2SR.

The architecture of the improved open-loop Swin2SR is shown in Fig. 1. It comprises three serial units: down-sampling (DS), compression (CP) and super-resolution (SR). The SR unit includes three subunits: shallow feature extraction (SF), deep feature extraction (DF) and reconstruction (RC). The DF subunit consists of multiple residual SwinV2 transformer blocks (RSTB) and a convolution block (CONV). $x_{h0}$ is the original high-resolution image, $x_{d0}$ is the down-sampling low-resolution image, $x_{c0}$ is the compressed low-resolution image, and $x_{s0}$ the reconstruction high-resolution image.

Compared with the classical Swin2SR, DS and CP units are extraordinarily added. The introduction of the DS unit is due to the reasonable assumption that a low-resolution image before compression can be regarded as a down-sampling version of a high-resolution image. Although the DS and CP units are implicitly utilized in the classical Swin2SR, they are not obviously analyzed. The DS unit can adopt any traditional down-sampling approaches, such as nearest neighbor and interpolation. The CP unit can employ any existing lossy image compression standards, such as JPEG and WebP. The DS and CP units are designed for training SR unit and evaluating SR recovery performance.

The improved Swin2SR can de described by following mathematical equations.

$$\mathbf{x}_{d0} = \mathrm{DS}(\mathbf{x}_{h0}), \mathbf{x}_{c0} = \mathrm{CP}(\mathbf{x}_{d0}), \mathbf{x}_{s0} = \mathrm{SR}(\mathbf{x}_{c0})$$
$$\mathbf{x}_{s0} = \mathrm{SR}\left(\mathrm{CP}\left(\mathrm{DS}(\mathbf{x}_{h0})\right)\right) \quad , \quad (1)$$
$$\mathbf{x}_{h0}, \mathbf{x}_{s0} \in \mathrm{R}^D; \mathbf{x}_{d0}, \mathbf{x}_{c0} \in \mathrm{R}^d$$

where: D is the dimension of high-resolution image and d is the dimension of low-resolution image.

### B. Proposed Closed-Loop CSwin2SR

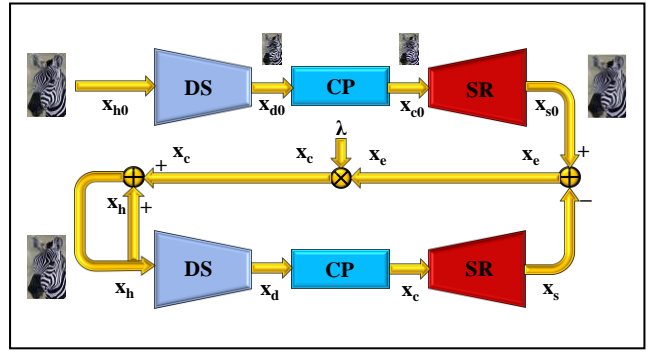

Fig. 2. Framework of proposed closed-loop CSwin2SR.

The framework of the proposed closed-loop CSwin2SR is displayed in Fig.2. It is composed of two elements: the top half is a serial Swin2SR and the bottom half is a circular Swin2SR. The serial Swin2SR obtains the initial high-quality reconstruction $x_{s0}$; the circular Swin2SR obtains the enhanced high-quality reconstruction $x_h$.

The circular Swin2SR comprises six modules: DS, CP, SR, two summators and multiplier. It constructs a closed-loop negative feedback system. $x_d$ is the down-sampling version of $x_h$, $x_c$ is the compression version of $x_d$, $x_r$ the reconstruction version of $x_c$. $x_e$ is the error vector and $x_c$ is the control vector.

The circular Swin2SR can be depicted by following mathematical expressions.

$$\mathbf{x}_d = \mathrm{DS}(\mathbf{x}_h), \mathbf{x}_c = \mathrm{CP}(\mathbf{x}_d), \mathbf{x}_s = \mathrm{SR}(\mathbf{x}_c)$$
$$\mathbf{x}_e = \mathbf{x}_{s0} - \mathbf{x}_s, \mathbf{x}_c = \lambda \mathbf{x}_e$$
$$\mathbf{x}_h \leftarrow \mathbf{x}_h + \mathbf{x}_c \quad , \quad (2)$$
$$\mathbf{x}_h \leftarrow \mathbf{x}_h + \lambda\left(\mathbf{x}_{s0} - \mathrm{SR}\left(\mathrm{CP}\left(\mathrm{DS}(\mathbf{x}_h)\right)\right)\right)$$
$$\mathbf{x}_h, \mathbf{x}_s, \mathbf{x}_e, \mathbf{x}_c \in \mathrm{R}^D; \mathbf{x}_d, \mathbf{x}_c \in \mathrm{R}^d; \lambda \in (0,1]$$

where: $\lambda$ is a small constant; symbol '←' means iterative updating; the initial value of $x_h$ can be any choices, such as zero vector, random vector or initial reconstruction $x_{s0}$.

According to the theory of negative feedback, the error vector is close to zero when the circular Swin2SR system reaches steady state. Therefore, $x_s$ approaches $x_{s0}$ and $x_h$ approaches $x_{h0}$. Ultimately, the perfect high-resolution reconstruction $x_h$ is attained. The whole procedure can be expressed by the following mathematical formulas.

$$\mathbf{x}_e \to 0$$
$$\mathbf{x}_e = \mathbf{x}_{s0} - \mathbf{x}_s \Rightarrow \mathbf{x}_s \to \mathbf{x}_{s0}$$
$$\mathbf{x}_s = \mathrm{SR}\left(\mathrm{CP}\left(\mathrm{DS}(\mathbf{x}_h)\right)\right), \mathbf{x}_{s0} = \mathrm{SR}\left(\mathrm{CP}\left(\mathrm{DS}(\mathbf{x}_{h0})\right)\right), \quad (3)$$
$$\Rightarrow \mathbf{x}_h \to \mathbf{x}_{h0}$$

The proposed CSwin2SR algorithm can be described by following Fig. 3.

```
Algorithm: CSwin2SR
Input: x_c0
Initialization:
    λ = 0.1, n = 1, x_s0 = SR(x_c0), x_h = x_s0
while n <= 10
    x_h = x_h + λ(x_s0 − SR(CP(DS(x_h))))
    n = n+1
end
Output: x_h
```

Fig. 3. The proposed CSwin2SR algorithm.

## IV. EXPERIMENTS

### A. Experimental Conditions

The experimental section is designed to evaluate the reconstruction capability of the proposed method. The experimental hardware platform is NVIDIA GPUs performing on the cloud and the experimental software platform is Pytorch running on Linux operating system. The open-source codes and corresponding pretrained models of Swin2SR is employed (https://github.com/mv-lab/swin2sr). The DIV2K valid dataset and Set5 dataset are utilized; the former holds 100 high-resolution colorful images with size form 1320×2040 to 1368×2040 and the latter holds 5 colorful images with size form 228×344 to 512×512 [10]. The lossy image compression algorithm is JPEG. The performance metrics include peak signal-to-noise ratio (PSNR) and structural similarity (SSIM). For the purpose of computing PSNR and SSIM, the low-resolution image is up-sampled to the same size as the original image. The experimental parameters are listed in Tab. 1.

TABLE I. EXPERIMENTAL PARAMETERS

| Symbol | Name | Value |
|---|---|---|
| λ | Iterative constant | 0.1 |
| N | Total number of iteration | 10 |
| q | JPEG compression quality | 10% |
| dr | Downsampling rate | 4 |
| sr | Super-resolution factor | 4 |

### B. Experimental Results

Tab. 2 and 3 summarize the average PSNR and SSIM on DIV2K valid dataset and Set5 dataset. The increment of average PSNR is larger than 0.18 dB and the related increment of average SSIM is larger than 0.01. Compared with classical Swin2SR, the proposed CSwin2SR efficiently and hugely improves the capability of super-resolution reconstruction.

Fig. 4, 5 and 6 illustrate exemplar images for comparing the proposed CSwin2SR and the classical Swin2SR, where subfigure (1) is the original high-resolution image; subfigure (2) is the compressed low-resolution image of JPEG which is zoomed up to the same size as the original image; subfigure (3) is the high-resolution reconstruction image of classical Swin2SR; subfigure (4) is the enhanced high-resolution reconstruction image of propose CSwin2SR; subfigure (5) is the enlarged image patch which is marked with red box in the original image; subfigure (6) is the enlarged image patch of the compressed image; subfigure (7) is the enlarged image patch of Swin2SR reconstruction image; subfigure (8) is the enlarged image patch of CSwin2SR reconstruction image; subfigure (9) is the image patch of absolute difference between subfigure (7) and subfigure (5); subfigure (10) is the image patch of absolute difference between subfigure (8) and subfigure (5). Because it is uneasy for human visual system to discriminate the reconstruction image patches of Swin2SR and CSwin2SR, the image patch of absolute difference is produced. The absolute difference demonstrates that the proposed CSwin2SR has smaller reconstruction error than the classical Swin2SR. In a word, the proposed CSwin2SR is superior to the classical Swin2SR in the performance of SR reconstruction.

According to subfigure (9) and (10) in Fig. 4, 5 and 6, the lossy JPEG compression results in horizontal, vertical and diagonal shifts of pixels. The proposed CSwin2R can more efficiently eliminate the shifts than the traditional Swin2SR.

It should be remarkable that the PSNR increment of image patch in subfigure (8) of Fig. 4 compared with that of subfigure (7) is greater than 4dB and the related SSIM increment is greater than 0.01. The PSNR increment of image patch in subfigure (8) of Fig. 5 compared with that of subfigure (7) is close to 2dB and the related SSIM increment is close to 0.03. The PSNR increment of image patch in subfigure (8) of Fig. 6 compared with that of subfigure (7) is greater than 2dB and the related SSIM increment is greater than 0.06. It powerfully demonstrates the outstanding image SR reconstruction capability of the proposed CSwin2SR.

TABLE II. EXPERIMENTAL RESULTS ON DIV2K VALID DATASET

| DS | Method | Performance Metrics | |
|---|---|---|---|
| | | PSNR(dB) | SSIM |
| bicubic | Swin2SR | 23.59 | 0.6409 |
| | CSwin2SR | 23.60 | 0.6410 |
| | Increment | 0.01 | 0.0001 |
| nearest | Swin2SR | 21.96 | 0.6165 |
| | CSwin2SR | 22.14 | 0.6207 |
| | Increment | 0.18 | 0.0105 |

TABLE III. EXPERIMENTAL RESULTS ON SET5 DATASET

| DS | Method | Performance Metrics | |
|---|---|---|---|
| | | PSNR(dB) | SSIM |
| bicubic | Swin2SR | 22.28 | 0.6315 |
| | CSwin2SR | 22.29 | 0.6316 |
| | Increment | 0.01 | 0.0001 |
| nearest | Swin2SR | 20.79 | 0.5901 |
| | CSwin2SR | 21.07 | 0.5972 |
| | Increment | 0.28 | 0.0061 |

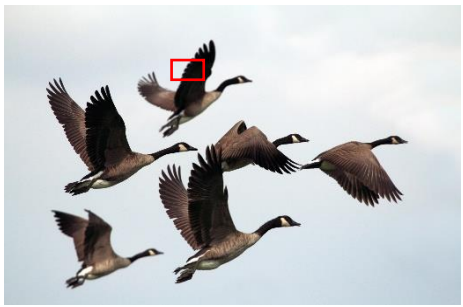
(1) Original

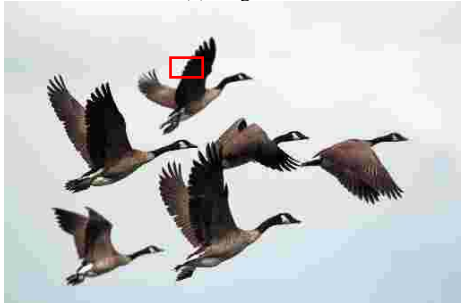
(2) Compressed (PSNR=25.80, SSIM=0.7663)

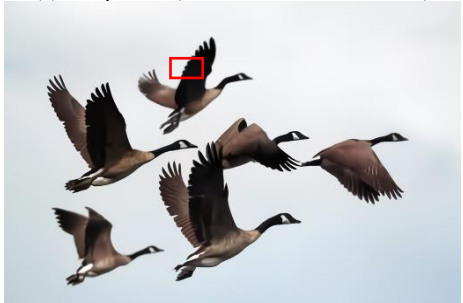
(3) Swin2SR (PSNR=26.83, SSIM=0.8901)

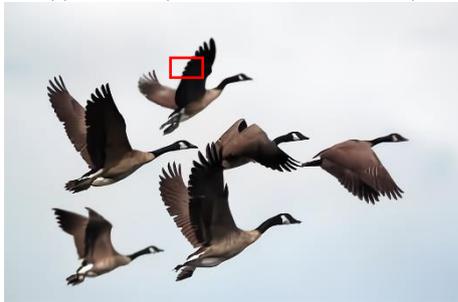
(4) CSwin2SR (PSNR=28.28, SSIM=0.8920)

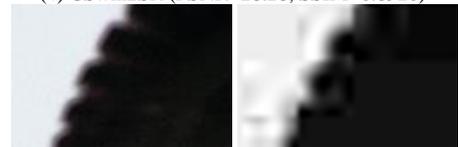
(5) Original  (6) Compressed (PSNR=24.59, SSIM=0.7414)

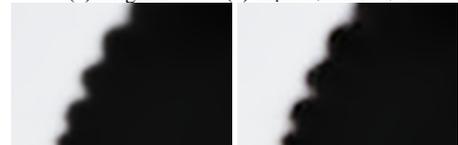
(7) Swin2SR (PSNR=25.29, SSIM=0.8642)  (8) CSwin2SR (PSNR=29.98, SSIM=0.8749)

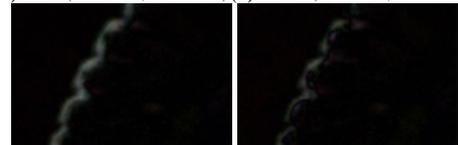
(9) Swin2SR  (10) CSwin2SR

Fig. 4. Comparison between Swin2SR and CSwin2SR (goose).

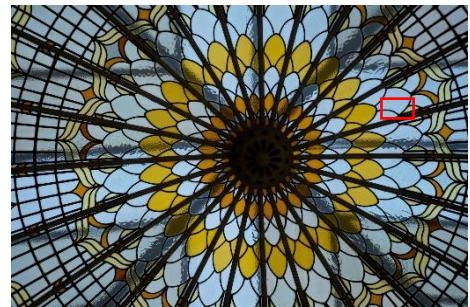
(1) Original

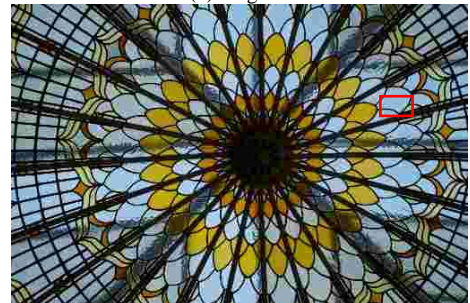
(2) Compressed (PSNR=17.76, SSIM=0.5709)

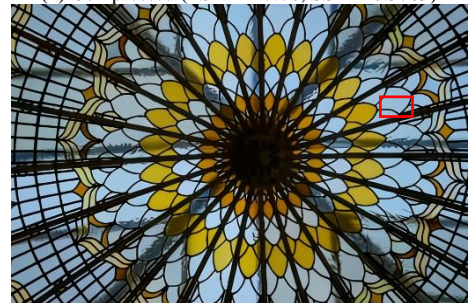
(3) Swin2SR (PSNR=18.51, SSIM= 0.7045)

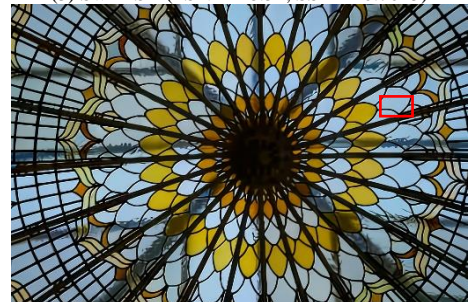
(4) CSwin2SR (PSNR=19.71, SSIM=0.7355)

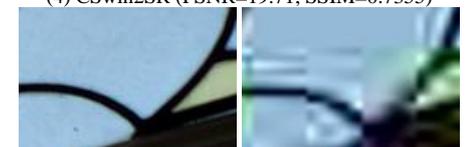
(5) Original  (6) Compressed (PSNR=18.72, SSIM=0.7571)

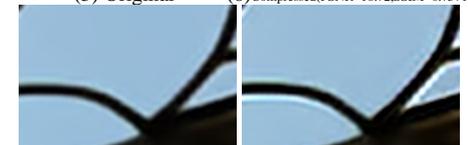
(7) Swin2SR (PSNR=18.96, SSIM=0.8081)  (8) CSwin2SR (PSNR=20.96, SSIM=0.8389)

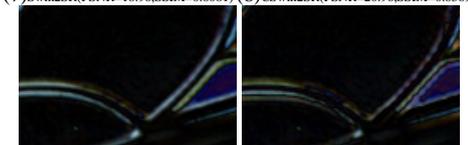
(9) Swin2SR  (10) CSwin2SR

Fig. 5. Comparison between Swin2SR and CSwin2SR (roof).

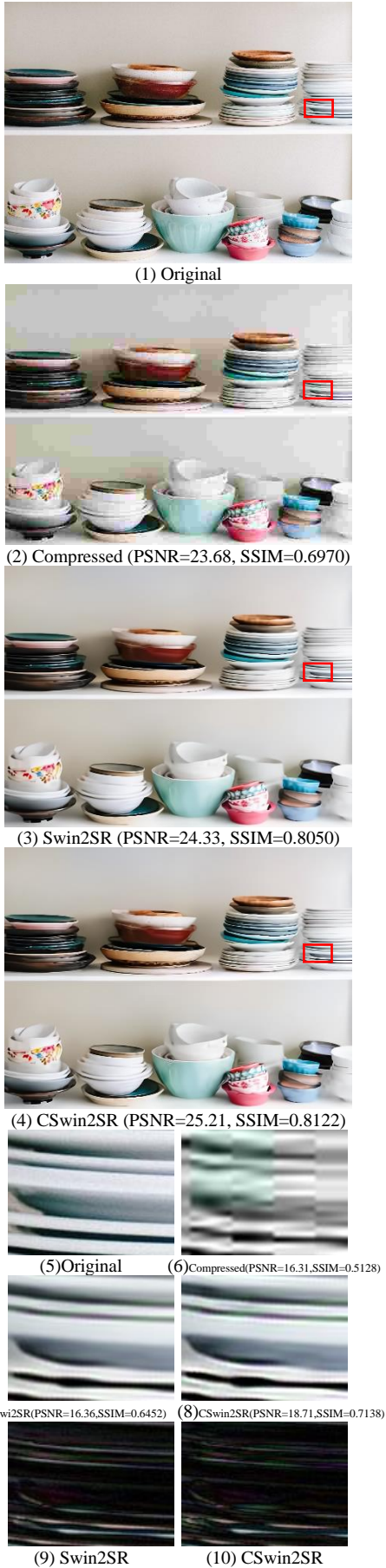

(1) Original

(2) Compressed (PSNR=23.68, SSIM=0.6970)

(3) Swin2SR (PSNR=24.33, SSIM=0.8050)

(4) CSwin2SR (PSNR=25.21, SSIM=0.8122)

(5) Original  (6) Compressed(PSNR=16.31,SSIM=0.5128)

(7) Swi2SR(PSNR=16.36,SSIM=0.6452)  (8) CSwin2SR(PSNR=18.71,SSIM=0.7138)

(9) Swin2SR  (10) CSwin2SR

Fig. 6. Comparison between Swin2SR and CSwin2SR (cupboard).

## V. CONCLUSIONS

This paper presents a closed-loop CSwin2R method for compressed image SR. It consist of a serial Swin2SR and a circular Swin2SR. The serial Swin2SR is an improvement of classical Swin2SR by supplementing a down-sampling unit and a compression unit. The circular Swin2SR is a closed-loop architecture by introducing a negative feedback mechanism. It is verified by the experimental results that the proposed CSwin2SR is superior to the classical Swin2SR in the capability of image SR reconstruction. The proposed CSwin2SR can efficiently remove pixel shifts in the reconstruction image of the classical Swin2SR. On DIV2K test and valid datasets, the increment of average PSNR is greater than 0.18 dB and the related increment of SSIM is greater than 0.01.

In our future work, the theory basis of CSwin2SR will be deeply studied. More SR factors, compression algorithms, image datasets will also be explored.

## ACKNOWLEDGMENT

We would very much like to thank the authors of Swin2SR for selflessly sharing their open-source codes on Microsoft GitHub. We also express deep thanks to Google COLAB for free GPU computing service.